\documentclass{amsart}
\usepackage{amsthm,amsmath,amssymb,amsfonts}
\title{The Promise Polynomial Hierarchy}
\author[Adam Chalcraft]{Adam Chalcraft}
\address[Adam Chalcraft]{IDA/CCR-L, 4320 Westerra Court, San Diego, CA 92121}
\email[Corresponding author]{adam.chalcraft@gmail.com}
\author[Samuel Kutin]{Samuel Kutin}
\address[Samuel Kutin]{IDA/CCR-P, 805 Bunn Drive, Priceton NJ 08540}
\email{kutin@idaccr.org}
\author[David Petrie Moulton]{David Petrie Moulton}
\address[David Petrie Moulton]{IDA/CCR-P, 805 Bunn Drive, Priceton NJ 08540}
\email{moulton@idaccr.org}
  
\theoremstyle{plain}
\newtheorem{lemma}{Lemma}
\newtheorem{theorem}[lemma]{Theorem}

\parskip=\smallskipamount
\def\bigand{\bigwedge}
\def\bigor{\bigvee}
\def\card#1{\big|#1\big|}
\def\defn#1{{\sl #1}}
\def\implies{\Rightarrow}
\def\intersect{\mathop{\cap}}
\def\max{{\mathrm{max}}}
\def\min{{\mathrm{min}}}
\def\otherwise{\mathrm{otherwise}}
\def\squish#1{\makebox[0pt][c]{\hss$#1$\hss}}
\def\narr{\squish{\nearrow}}
\def\sarr{\squish{\searrow}}
\def\arr{\squish{\to}}
\def\xarr{\narr\sarr}

\def\A{\Sigma}
\def\PA{\mathbb P_+\A}
\def\promise#1#2{\left\{\begin{matrix}#1&(#2)\\
\A&\otherwise\end{matrix}\right.}
\def\first{\pi_1!}
\def\True{\textsf{True}}
\def\False{\textsf{False}}
\def\Binary{\{\True,\False\}}
\def\SAT{{\mathrm{SAT}}}
\def\CSAT{{\overline{\SAT}}}
\def\MVAL{{\mathrm{MaxVAL}}}
\def\mVAL{{\mathrm{MinVAL}}}
\def\VAL{{\mathrm{VAL}}}
\def\USAT{{\mathrm{USAT}}}
\def\UCSAT{{\overline{\USAT}}}
\def\UVAL{{\mathrm{UVAL}}}
\def\uSAT{{\mathrm{uSAT}}}
\def\xSAT{{\mathrm{xSAT}}}
\def\P{\widehat{\mathrm{P}}}
\def\S{\widehat\Sigma}
\def\CS{\widehat\Pi}
\def\V{\widehat{\mathrm{V}}}
\def\D{\widehat\Delta}
\def\U{\mathrm{U}}
\def\US{\U\S}
\def\UCS{\U\CS}
\def\UV{\U\V}
\def\UD{\U\widehat\Delta}
\def\PP{{\mathrm{P}}}
\def\NP{{\mathrm{NP}}}
\def\coNP{{\mathrm{coNP}}}
\def\UP{{\mathrm{UP}}}
\def\coUP{{\mathrm{coUP}}}
\def\BPP{{\mathrm{BP}\P}}
\def\BPPP{{\mathrm{BP}\PP}}

\begin{document}
\begin{abstract}
The polynomial hierarchy is a grading of problems by difficulty,
including $\PP$, $\NP$ and $\coNP$ as the best known classes.  The
\defn{promise} polynomial hierarchy is similar, but extended to
include promise problems.  It turns out that the promise polynomial
hierarchy is considerably simpler to work with, and many open
questions about the polynomial hierarchy can be resolved in the
promise polynomial hierarchy.

Goldreich \cite{Goldreich} argues that promise problems are a more
natural object of study than non-promise problems, and our results
would seem to confirm this.

Our main theorem is that, in the world of promise problems, if
$\phi\propto_T\SAT$ then $\phi\propto\UVAL_2$, where $\UVAL_2(f)$ is
the promise problem of finding the unique $x$ such that $\forall
y,\,f(x,y)=1$.  We also give a complete promise problem for the
promise problem equivalent of $\UP\intersect\coUP$, and prove the
promise problem equivalents of $\PP^{\UP\intersect\coUP}=\PP^\UP$ and
$\BPPP^{\UP\intersect\coUP}=\BPPP^\NP$.  Analagous results are known
for $\NP$ and $\coNP$ \cite{ESY}.
\end{abstract}

\maketitle

\section{Definitions}
\subsection{The alphabet}
We use the alphabet $\A=\{0,1\}$.  If $x\in\A^n$, we write $|x|=n$ and
$x=(x_1,\ldots,x_n)$.  If $1\leq i\leq n$, we write $\pi_i(x)=x_i$.
If $x\in\A^n$, we write $\lnot x=(\lnot x_1,\ldots,\lnot x_n)$.  We
regard $\A^n$ as totally ordered lexicographically, with $x_1$ as the
most important, so $(0,1)<(1,0)$.  We write
$\A^*=\bigcup_{n=1}^\infty\A^n$ for the set of finite strings from
$\A$.  We write $\PA=\{\{0\},\{1\},\A\}$ for the set of non-empty
subsets of $\A$.  We define $\lnot:\PA\to\PA$ by $\lnot\{0\}=\{1\}$,
$\lnot\{1\}=\{0\}$ and $\lnot\A=\A$.

\subsection{Promise problems}
We extend the usual definition of a \defn{problem} to a \defn{promise
problem}.  Promise problems were first introduced by Even, Selman and
Yacobi \cite{ESY}.  The concept of a promise problem encompasses two
generalizations of a problem simultaneously.  Firstly, a promise
problem comes with a promise which we may assume is satisfied; if the
promise is not satisfied, then either answer is valid.  Secondly, a
promise problem allows for more than one valid answer to a problem.

Our approach is to reduce non-promise problems to promise problems and
then work exclusively with promise problems in the sequel.

A \defn{problem} (or, for emphasis, a \defn{non-promise} problem) is a
function $\phi:\A^*\to\A$.  A function $\sigma:\A^*\to\A$
\defn{solves} $\phi$ when $\sigma=\phi$.  Let $\Phi$ be the set of all
non-promise problems.

A \defn{promise problem} is a function $\phi:\A^*\to\PA$. A function
$\sigma:\A^*\to\A$ \defn{solves} $\phi$ when $\forall
x\in\A^*,\,\sigma(x)\in\phi(x)$.  We will treat a problem $\phi$ as a
promise problem $\hat\phi$ by defining $\hat\phi(x)=\{\phi(x)\}$.

There are many other possible equivalent definitions of a promise
problem \cite{ESY, Goldreich}.  One is to define a promise problem as
a pair $(\phi,\psi)$, where $\phi:\A^*\to\A$ is the problem and
$\psi:\A^*\to\Binary$ is the promise.  Under this definition, a
function $\sigma:\A^*\to\A$ solves $(\phi,\psi)$ when $\forall
x\in\A^*,\,(\psi(x)\implies\sigma(x)=\phi(x))$.  We treat such a pair
as a promise problem $\hat\phi$ by defining $\hat\phi(x)=\{\phi(x)\}$
when $\psi(x)=\True$ and $\hat\phi(x)=\A$ when $\psi(x)=\False$.

\subsection{Strong reduction}
If $\phi_1$ and $\phi_2$ are promise problems, we write
$\phi_1\supseteq\phi_2$ to mean that, for all $x\in\A^*$,
$\phi_1(x)\supseteq\phi_2(x)$.  Equivalently, if $\sigma$ solves
$\phi_2$ then $\sigma$ solves $\phi_1$.  This equivalence requires
$\emptyset\notin\PA$, which is why we defined $\PA$ as we did.
Informally, $\phi_1$ is ``easier'' than (or as easy as) $\phi_2$.

If $\phi_1$ and $\phi_2$ are non-promise problems, we remind the
reader that $\phi_1\propto\phi_2$ means that there is a Turing machine
$M$ which runs in time polynomial in $|x|$ and computes some function
$\mu:\A^*\to\A^*$ such that $\forall
x\in\A^*,\,\phi_1(x)=\phi_2(\mu(x))$.  This is sometimes called
\defn{Karp reduction} or \defn{many-one reduction}.

If $\phi_1$ and $\phi_2$ are promise problems, we define
$\phi_1\propto\phi_2$ to mean that there is a Turing machine $M$ which
runs in time polynomial in $|x|$ and computes some function
$\mu:\A^*\to\A^*$ such that, for all $x\in\A^*$,
$\phi_1(x)\supseteq\phi_2(\mu(x))$.  Note that this generalization of
$\propto$ to promise problems is still transitive.

For a promise problem $\phi_2$, we define the \defn{strong closure}
$[\phi_2]=\{\phi_1\mid\phi_1\propto\phi_2\}$.  For example,
$[\SAT]\intersect\Phi=\NP$.  Note that $[\phi]$ is the
downward-closure of an equivalence class under $\propto$.

\subsection{Weak reduction}
If $\phi$ is a non-promise problem, we remind the reader that a
$\phi$-\defn{oracle} Turing machine is a Turing machine with the extra
ability, given $x\in\A^*$, to compute $\phi(x)$ in unit time.

If $\phi$ is a promise problem, we define a $\phi$-\defn{oracle}
Turing machine to be a Turing machine $M$ with the extra ability,
given $x\in\A^*$, to compute some $y\in\phi(x)$ in unit time.  If
$\phi(x)=\A$, $M$ is non-deterministic as to the value of $y$.  In
particular, there is no requirement for two separate calls to the
oracle to return the same value of $y$.

If $\phi_1$ and $\phi_2$ are non-promise problems, we remind the
reader that $\phi_1\propto_T\phi_2$ means that there is a
$\phi_2$-oracle Turing machine $M$ which runs in time polynomial in
$|x|$ and computes $\phi_1(x)$.  This is sometimes called \defn{Turing
reduction} or \defn{Cook reduction}.

If $\phi_1$ and $\phi_2$ are promise problems, we define
$\phi_1\propto_T\phi_2$ to mean that there is a $\phi_2$-oracle Turing
machine $M$ and a polynomial $p$ such that, when $M$ is given
$x\in\A^*$, every possible path of $M$ runs in time at most $p(|x|)$
and computes some $y\in\phi_1(x)$.  If $\phi_1(x)=\A$, there is no
need for different paths to return the same value of $y$.  Note that
this generalization of $\propto_T$ to promise problems is still
transitive.

Intuitively, $M$ must compute some $y\in\phi_1(x)$ in polynomial time
even if an adversary is watching the computation and choosing the
values returned by the $\phi_2$-oracle when there is a choice.

The condition $\phi_1\propto_T\phi_2$ is weaker than
$\phi\propto\phi_2$, because $M$ can call the $\phi_2$-oracle
polynomially many times.  In contrast, $\phi\propto\phi_2$ means that
(i) $M$ may only call the oracle for $\phi_2$ once, and (ii) it must
return the output from this oracle call unchanged.

For a promise problem $\phi_2$, we define the \defn{weak closure}
$\P^{\phi_2}=\{\phi_1\mid\phi_1\propto_T\phi_2\}$.  Note that
$\P^{\phi}$ is the downward-closure of an equivalence class under
$\propto_T$.  Since strong reduction implies weak reduction,
$\P^{\phi}$ is a union of strong equivalence classes, so we can define
$\P^{[\phi]}=\P^{\phi}$.

\subsection{Duality}
If $\phi$ is a promise problem, we define the \defn{dual} promise
problem $\lnot\phi$ by $(\lnot\phi)(x)=\lnot(\phi(x))$.  Note that
$\lnot\lnot\phi=\phi$.  We write $\lnot[\phi]=[\lnot\phi]$.  For
example, $\lnot\NP=\coNP$.

\section{The promise polynomial hierarchy}
The promise polynomial heirarchy is the partial order on promise
problems given by strong reduction.  More accurately, it is what we
currently know about this partial order; it is of course possible that
more strong reduction is true than we can currently prove, for example
if $\PP=\NP$.

Our approach is to introduce promise problems gradually, in what we
hope is a natural order, and build the promise polynomial hierarchy as
we go.  We give frequent diagrams of the partial order as we progress.
We start by defining some complexity classes near the bottom of the
heirarchy.

\subsection{Level 1}
We define the following problems and promise problems.  In each case,
the input is an encoding of some function $f:\A^m\to\A$, and the
suppressed range of $x$ is $\A^m$.  The exact encoding is not
important, so long as it is not too inefficient.  A suitable encoding
would be the value of $m$ in binary and a boolean expression in the
variables $\{x_1,\ldots,x_m\}$ using the operators
$\{\land,\lor,\lnot\}$ and parentheses.

Firstly, the non-promise problems.  It is traditional to define $\SAT$
as taking an expression of the form $\bigand\left(\bigor
x_i\lor\bigor\lnot x_j\right)$, but allowing general expressions is
equivalent. See, for example, \cite{GJ} LO7 (Satisfiability of Boolean
Expressions).
\begin{gather}
\SAT(f)=1\iff\exists x:f(x)=1\\
\CSAT(f)=1\iff\forall x,\,f(x)=1
\end{gather}

Now the promise problems.  Note that the expression $\{\pi_1 x\mid
f(x)=1\}$ in the definition of $\VAL(f)$ can be equal to $\A$, whereas
the same expression in the definition of $\UVAL(f)$ cannot, because of
the condition.  Our $\VAL$ is similar to $\xSAT$ of Even, Selman and
Yacobi \cite{ESY} and our $\USAT$ is similar to $\uSAT$ of Goldreich
\cite{Goldreich}.  $[\USAT]$ is also known as promise-$\UP$ and
$[\UCSAT]$ is also known as promise-$\coUP$.
\begin{gather}
\MVAL(f)=\promise{\{\pi_1\max\{x\mid f(x)=1\}\}}{\exists x:f(x)=1}\\
\mVAL(f)=\promise{\{\pi_1\min\{x\mid f(x)=1\}\}}{\exists x:f(x)=1}\\
\VAL(f)=\promise{\{\pi_1 x\mid f(x)=1\}}{\exists x:f(x)=1}\\
\USAT(f)=\promise{\{\SAT(f)\}}{\card{\{x\mid f(x)=1\}}\leq 1}\\
\UCSAT(f)=\promise{\{\CSAT(f)\}}{\card{\{x\mid f(x)=0\}}\leq 1}\\
\UVAL(f)=\promise{\{\pi_1 x\mid f(x)=1\}}{\exists!x:f(x)=1}
\end{gather}

We treat these uniformly as promise problems as discussed earlier.
For example, in the proof of the following theorem, the equalities are
equalities of elements of $\PA$.

\begin{theorem}\label{thm:U1->S1}\
\begin{enumerate}
\item $\lnot[\SAT]=[\CSAT]$.
\item $\lnot[\MVAL]=[\mVAL]$.
\item $\lnot[\VAL]=[\VAL]$.
\item $\lnot[\USAT]=[\UCSAT]$.
\item $\lnot[\UVAL]=[\UVAL]$.
\item $[\SAT]=[\MVAL]$.
\item $\UVAL\propto\VAL\propto\MVAL$.
\item $\UVAL\propto\USAT\propto\SAT$.
\end{enumerate}
\end{theorem}

\proof\
\begin{enumerate}
\item $\lnot\SAT(f(x))=\CSAT(\lnot f(x))$.
\item $\lnot\MVAL(f(x))=\mVAL(f(\lnot x))$.
\item $\lnot\VAL(f(x))=\VAL(f(\lnot x))$.
\item $\lnot\USAT(f(x))=\UCSAT(\lnot f(x))$.
\item $\lnot\UVAL(f(x))=\UVAL(f(\lnot x))$.
\item\begin{itemize}
\item $\SAT(f(x))=\MVAL(f(x_2,\ldots,x_{m+1})\lor\lnot x_1)$.
\item $\MVAL(f(x))\supseteq\SAT(f(x_1,\ldots,x_m)\land x_1)$.
\end{itemize}
\item $\UVAL(f)\supseteq\VAL(f)\supseteq\MVAL(f)$.
\item\begin{itemize}
\item $\UVAL(f)\supseteq\USAT(f(1,x_2,\ldots,x_m))$.
\item $\USAT(f)\supseteq\SAT(f)$.
\endproof
\end{itemize}
\end{enumerate}

This theorem gives a partial order on promise problems by strong
reduction, which we summarize in the following diagram.  We write
$X\to Y$ to mean $X\subseteq Y$.  Duality is given by reflection in a
horizontal axis.
\begin{equation}
\begin{matrix}
&&[\USAT]&\arr&[\SAT]=[\MVAL]\\
&\narr&&\narr\\
[\UVAL]&\arr&[\VAL]\\
&\sarr&&\sarr\\
&&[\UCSAT]&\arr&[\CSAT]=[\mVAL]
\end{matrix}
\label{eq:partial1}
\end{equation}

\subsection{Level 2}
Level 2 of the hierarchy has a set of problems and promise problems
similar to those in level 1.  For the moment, however, we wish to
define just one of these, so that we can show how levels 1 and 2 are
related.  We define the following promise problem.  The input is an
encoding of some $f:\A^m\times\A^{m'}\to\A$, which we write as
$f(x,y)$, so the suppressed range of $x$ is $\A^m$ and the suppressed
range of $y$ is $\A^{m'}$.
\begin{align}
\UVAL_2(f)&=\promise
{\{\pi_1 x\mid\forall y,\,f(x,y)=1\}}{\exists!x:\forall y,\,f(x,y)=1}
\end{align}

\begin{theorem}\label{thm:S1->U2}\
\begin{enumerate}
\item $\lnot[\UVAL_2]=[\UVAL_2]$.
\item $\MVAL\propto\UVAL_2$.
\end{enumerate}
\end{theorem}

\proof For (i), $\lnot\UVAL_2(f(x,y))=\UVAL_2(f(\lnot x,y))$.

For (ii), suppose we are given $f(x):\A^m\to\A$.  Let $m'=m(m-1)/2$
and define $g(x,y):\A^m\times\A^{m'}\to\A$ as follows.
\begin{equation}
\begin{split}
g(x,y)=&f(x_1,\ldots,x_m)\\
&\land(x_1\lor\lnot f(1,y_{1,2},\ldots,y_{1,m}))\\
&\land(x_2\lor\lnot f(x_1,1,y_{2,3},\ldots,y_{2,m})\\
&\land\cdots\\
&\land(x_{m-1}\lor\lnot f(x_1,\ldots,x_{m-2},1,y_{m-1,m}))\\
&\land(x_m\lor\lnot f(x_1,\ldots,x_{m-1},1))
\end{split}
\end{equation}
Now $\MVAL(f)\supseteq\UVAL_2(g)$.  \endproof

We add this information to the partial order (\ref{eq:partial1}).
\begin{equation}
\begin{matrix}
&&[\USAT]&\arr&[\SAT]=[\MVAL]\\
&\narr&&\narr&&\sarr\\
[\UVAL]&\arr&[\VAL]&&&&[\UVAL_2]\\
&\sarr&&\sarr&&\narr\\
&&[\UCSAT]&\arr&[\CSAT]=[\mVAL]
\end{matrix}
\label{eq:easy1}
\end{equation}

Later (theorem \ref{thm:Pn->Un}), we will show that, in fact,
$[\SAT]\subseteq\P^{\SAT}\subseteq[\UVAL_2]$, but we use theorem
\ref{thm:S1->U2} in the proof of theorem \ref{thm:Pn->Un}, so we must
prove it independently.

\subsection{Level $n$}
For $n\geq 1$, we make the obvious definitions of $\SAT_n(f)$ and so
on, where we may suppress the subscript when $n=1$.  The input is an
encoding of some function $f:\A^{m_1}\times\cdots\times\A^{m_n}\to\A$,
which we write as $f(x,y^{(1)},\ldots,y^{(n-1)})$, so the suppressed
range of $x$ is $\A^{m_1}$ and the suppressed range of $y^{(i)}$ is
$\A^{m_{i+1}}$.  The function
$f_x:\A^{m_2}\times\cdots\times\A^{m_n}\to\A$ is the obvious function
obtained by fixing the value of $x$.

Firstly, the non-promise problems.
\begin{gather}
\SAT_n(f)=1\iff\exists x:\CSAT_{n-1}(f_x)=1\\
\CSAT_n(f)=1\iff\forall x,\,\SAT_{n-1}(f_x)=1
\end{gather}

Now the promise problems.
\begin{gather}
\MVAL_n(f)=\promise
{\{\pi_1\max\{x\mid\CSAT_{n-1}(f_x)=1\}\}}{\SAT_n(f)=1}\\
\mVAL_n(f)=\promise
{\{\pi_1\min\{x\mid\CSAT_{n-1}(f_x)=1\}\}}{\SAT_n(f)=1}\\
\VAL_n(f)=\promise
{\{\pi_1 x\mid\CSAT_{n-1}(f_x)=1\}}{\SAT_n(f)=1}\\
\USAT_n(f)=\promise{\{\SAT_n(f)\}}
{\card{\{x\mid\CSAT_{n-1}(f_x)=1\}}\leq 1}\\
\UCSAT_n(f)=\promise{\{\CSAT_n(f)\}}
{\card{\{x\mid\SAT_{n-1}(f_x)=0\}}\leq 1}\\
\UVAL_n(f)=\promise
{\{\pi_1 x\mid\CSAT_{n-1}(f_x)=1\}}{\exists!x:\CSAT_{n-1}(f_x)=1}
\end{gather}

For the following theorem, we hope that it is clear how to generalize
the proof from $n=1$ (theorems \ref{thm:U1->S1} and \ref{thm:S1->U2})
to general $n\geq 1$.  We feel that writing it out would require a
notation which adds nothing and obscures the ideas.

\begin{theorem}
For $n\geq 1$,
\begin{enumerate}
\item $\lnot[\SAT_n]=[\CSAT_n]$.
\item $\lnot[\MVAL_n]=[\mVAL_n]$.
\item $\lnot[\VAL_n]=[\VAL_n]$.
\item $\lnot[\USAT_n]=[\UCSAT_n]$.
\item $\lnot[\UVAL_n]=[\UVAL_n]$.
\item $[\SAT_n]=[\MVAL_n]$.
\item $\UVAL_n\propto\VAL_n\propto\MVAL_n\propto\UVAL_{n+1}$.
\item $\UVAL_n\propto\USAT_n\propto\SAT_n$.
\end{enumerate}
\end{theorem}

\proof Omitted.  \endproof

We add this information to the partial order (\ref{eq:easy1}).  To
save space, we use the following notation, where $\S_n$ and $\CS_n$
are chosen to be similar to the standard notation (see section
\ref{sec:standard} below).
\begin{align}
\S_n&=[\SAT_n]=[\MVAL_n]&\US_n&=[\USAT_n]\\
\CS_n&=[\CSAT_n]=[\mVAL_n]&\UCS_n&=[\UCSAT_n]\\
\V_n&=[\VAL_n]&\UV_n&=[\UVAL_n]
\end{align}
\begin{equation}
\begin{matrix}
&&\US_1&\arr&\S_1&&&&\US_2&\arr&\S_2\\
&\narr&&\narr&&\sarr&&\narr&&\narr&&\sarr&&\narr\\
\UV_1&\arr&\V_1&&&&\UV_2&\arr&\V_2&&&&\UV_3&\arr\\
&\sarr&&\sarr&&\narr&&\sarr&&\sarr&&\narr&&\sarr\\
&&\UCS_1&\arr&\CS_1&&&&\UCS_2&\arr&\CS_2
\end{matrix}
\label{eq:easyn}
\end{equation}

\subsection{Level 0}
We have so far ignored the very bottom of the hierarchy.  Let $\P$ be
the set of promise problems $\phi$ such that there is a
polynomial-time Turing machine $M$ which computes some function
$\mu:\A^*\to\A$ which solves $\phi$.  Alternatively, $\P=\P^0$, where
$0$ is the trivial problem $0(x)=0$ for all $x\in\A^*$.  The next
theorem is nearly trivial.

\begin{theorem}
$\P\subseteq\UV_1$.
\end{theorem}

\proof Clearly $\pi_1\in\P$.  Let $\phi\in\P$ and $\mu$ be as above.
Now $\pi_1(\mu(x))=\mu(x)$, so $\phi\propto\pi_1$, so $\P=[\pi_1]$.
Now $\pi_1(x)=\UVAL(f_x)$, where $f_x(y)=1\iff y=(x_1)$.  \endproof

\subsection{The standard hierarchy}
\label{sec:standard}
For $n\geq 1$, let $\D_{n+1}=\P^{\SAT_n}$.  We remind the reader that
$\Phi$ is the set of non-promise problems.  We define
$\PP=\P\intersect\Phi$, $\Sigma_n=\S_n\intersect\Phi$,
$\Pi_n=\CS_n\intersect\Phi$, and $\Delta_n=\D_n\intersect\Phi$ as
usual.  We have $\Sigma_1=\NP$ and $\Pi_1=\coNP$.  We remind the
reader of the standard polynomial hierarchy \cite{MS, GJ}.
\begin{equation}
\begin{matrix}
&&\Sigma_1&&&&\Sigma_2\\
&\narr&&\sarr&&\narr&&\sarr&&\narr\\
\PP&&&&\Delta_2&&&&\Delta_3\\
&\sarr&&\narr&&\sarr&&\narr&&\sarr\\
&&\Pi_1&&&&\Pi_2
\end{matrix}
\label{eq:standard}
\end{equation}

We now show that this partial order remains true if we include promise
problems, using $\P$, $\S_n$, $\CS_n$ and $\D_n$.  We already know
that $\P\subseteq\UV_1\subseteq\S_1$.

\begin{theorem}
$\S_n\subseteq\D_{n+1}$.
\end{theorem}

\proof $\SAT_n\in\P^{\SAT_n}$.  \endproof

The proof of the following theorem is the standard argument which
shows that $\Delta_2\subseteq\Sigma_2$.  We write it out in detail,
partly so that we can check that it still works for promise problems,
but mainly because we will need all the detail anyway when we come to
prove theorem \ref{thm:Pn->Un}, and it helps to see it in a more
familiar context first.

Incidentally, one might think that the standard argument which shows
that $\Delta_2\subseteq\Sigma_2$ is just $\PP\subseteq\NP$ relative to
any oracle, so $\PP^\NP\subseteq\NP^\NP$, so
$\Delta_2\subseteq\Sigma_2$.  This argument is fine, but then one has
to show that $\SAT_2$ is a complete problem for $\NP^\NP$, which is
essentially the same theorem.

\begin{theorem}\label{thm:Pn->Sn}
For $n\geq 2$, $\D_n\subseteq\S_n$.
\end{theorem}

\proof We give the proof for $n=2$.

Suppose $\phi\in\D_2=\P^{\SAT}$.  Then $\phi$ can be solved by
a Turing machine $M$ which takes $\alpha\in\A^*$ as input, runs in
time at most $T$, where $T$ is polynomial in $m=|\alpha|$, and which
is allowed to make calls to a $\SAT$ oracle.  Conveniently, $\SAT$ is
a non-promise problem, so $M$ is deterministic, so we can say that $M$
calculates $\sigma$, where $\sigma$ solves $\phi$.  Let $X_{i,t}$ be
the value on the tape at position $i$ and time $t$ and let $Y_{s,t}$
be 1 if $M$ is in state $s$ at time $t$ and 0 otherwise.  The ranges
are $-T\leq i\leq T$ and $0\leq t\leq T$, because a Turing machine
moves at most 1 cell at each step.

This means that $\sigma(\alpha)$ can be written as a circuit $f$ with
$n$ gates, where $n$ is polynomial is $m$, but with the ability to use
a special $\SAT$ gate, which evaluates $\SAT(g)$, where $g$ is (an
encoding of) a circuit which is the input to the gate.  Let the values
at the vertices of $f$ be $\alpha=(\alpha_1,\ldots,\alpha_m)$ and
$x=(x_1,\ldots,x_n)$, and label these vertices (conveniently, but
rather unusually) such that $x_1$ is the output vertex.  We can write
the condition $v(x)$ that $x$ is a valid assignment of values to the
vertices as follows, where $a_i(\alpha,x)$ is a function representing
the gate which calculates the value at the vertex $x_i$.
\begin{equation}
\begin{split}
v(\alpha,x)&=\bigand_{i=1}^n(x_i=a_i(\alpha,x))\\
&=\bigand_{i=1}^n
(\lnot x_i\lor a_i(\alpha,x))\land(x_i\lor\lnot a_i(\alpha,x))
\end{split}
\end{equation}

For an ordinary gate, $a_i(\alpha,x)$ is a polynomial expression.  For
a $\SAT$ gate, we can write $a_i(\alpha,x)=\exists y:b_i(\alpha,x,y)$,
where $b_i(\alpha,x,y)$ is a polynomial expression.  We can therefore
write $v(\alpha,x)$ as follows, for some polynomial expressions $p_1$
and $p_2$.
\begin{equation}
v(\alpha,x)=
(\exists y:p_1(\alpha,x,y))\land(\forall z,\,p_2(\alpha,x,z))
\end{equation}

We write $\sigma(\alpha)=1\iff\exists x:v(\alpha,x)\land(x_1=1)$,
which simplifies to $\exists(x,y):\forall z,\,p_3(\alpha,x,y,z)$ for
some polynomial expression $p_3$.  Now we write
$p_3(\alpha,x,y,z)=h_\alpha(x,y,z)$ for some circuit $h_\alpha$ of
polynomial size which we can construct from $\alpha$ in polynomial
time.  We therefore have $\sigma(\alpha)=1\iff\exists(x,y):\forall
z,\,h_\alpha(x,y,z)$, so $\sigma\in\Sigma_2$.

Now $\phi\propto\sigma\in\Sigma_2\subseteq\S_2$, so
$\phi\in[\sigma]\subseteq\S_2$ (because $\S_2$ is a union of
strong equivalence classes) as required.  \endproof

We therefore have the promise version of the standard polynomial
hierarchy (\ref{eq:standard}).
\begin{equation}
\begin{matrix}
&&\S_1&&&&\S_2\\
&\narr&&\sarr&&\narr&&\sarr&&\narr\\
\P&&&&\D_2&&&&\D_3\\
&\sarr&&\narr&&\sarr&&\narr&&\sarr\\
&&\CS_1&&&&\CS_2
\end{matrix}
\label{eq:standardhat}
\end{equation}

\subsection{Combining the hierarchies}
This raises the natural question of how the partial orders
(\ref{eq:easyn}) and (\ref{eq:standardhat}) fit together.  We already
know that $\P\subseteq\UV_1$.

For the next theorem, we need a notation.  Suppose $X$ is a set,
suppose $P:X\to\Binary$ is some predicate, and suppose $\exists!x\in
X:P(x)$.  Then we define $!x:P(x)$ to be that element.  In other
words, $P(!x:P(x))$ is true.  If $\exists!x\in X:f(x)=1$, then
$\UVAL(f)=\{\first x:f(x)=1\}$.

\begin{theorem}\label{thm:Pn->Un}
For $n\geq 2$, $\D_n\subseteq\UV_n$.
\end{theorem}

\proof We give the proof for $n=2$.  We use the notation and most of
the proof of theorem \ref{thm:Pn->Sn}.

The point is that there is exactly one assignment of $x$ which
satisfies $v(\alpha,x)$, and $x_1$ is the output vertex, so we can
write $\sigma(\alpha)=\first x:v(\alpha,x)$.
\begin{equation}
\sigma(\alpha)=\first x:
(\exists y:p_1(\alpha,x,y))\land(\forall z,\,p_2(\alpha,x,z))
\end{equation}

Now theorems \ref{thm:U1->S1} and \ref{thm:S1->U2} say that
$\SAT\propto\UVAL_2$ and, by duality, $\CSAT\propto\UVAL_2$.  This
means that we can transform the expressions $\exists
y:p_1(\alpha,x,y)$ and $\forall z,\,p_2(\alpha,x,z)$ into $\UVAL_2$
form in polynomial time, and therefore polynomially many extra
variables.
\begin{align}
\exists y:p_1(\alpha,x,y)&\iff
(\first s:\forall z,\,p_3(\alpha,s,x,z))=1\\
\forall z,\,p_2(\alpha,x,z)&\iff
(\first t:\forall z,\,p_4(\alpha,t,x,z))=1
\end{align}

Note that, for all $x\in\A^*$, it is true both that $\exists!s:\forall
z,\,p_3(\alpha,s,x,z)$ and that $\exists!t:\forall z,\,p_4(\alpha,t,x,z)$,
so the $!s:\ldots$ notation is justified.  We substitute these into
$\sigma(\alpha)$.
\begin{equation}
\sigma(\alpha)=\first x:
(\first s:\forall z,\,p_3(\alpha,s,x,z))
(\first t:\forall z,\,p_4(\alpha,t,x,z))=1
\end{equation}

We repeatedly simplify this expression for $\sigma(\alpha)$.  We
define $u=(s_1 t_1)$ and we define $v=(u,s,t)$ and $w=(x,v)$ to be the
simple concatenation of previous variables.  Note that the order is
important, so that $v_1=u_1=s_1 t_1$ and $w_1=x_1$.  We also define
new polynomial expressions $p_i$ as we go.
\begin{align}
&\sigma(\alpha)=\ldots\notag\\
&\first x:(\first(u,s,t):(u=s_1 t_1)\land
\forall z,\,p_3(\alpha,s,x,z)\land p_4(\alpha,t,x,z))=1\\
&\first x:(\first v:\forall z,\,p_5(\alpha,v,x,z))=1\\
&\first(x,v):(v_1=1)\land\forall z,\,p_5(\alpha,v,x,z)\\
&\first w:\forall z,\,p_6(\alpha,w,z)
\end{align}

Now we can write $p_6(\alpha,w,z)=h_\alpha(w,z)$, for some circuit
$h_\alpha$ of polynomial size which we can construct from $\alpha$ in
polynomial time.  We therefore have
$\sigma(\alpha)=\first w:\forall z,\,h_\alpha(w,z)$, so $\sigma\in
\UV_2$.

Now $\phi\propto\sigma\in\UV_2$, so $\phi\in[\sigma]\subseteq\UV_2$ as
required.  \endproof

We can now combine the partial orders (\ref{eq:easyn}) and
(\ref{eq:standardhat}).
\begin{equation}
\begin{matrix}
&&\US_1&\arr&\S_1&&&&\US_2&\arr&\S_2\\
&\narr&&\narr&&\sarr&&\narr&&\narr&&\sarr\\
\P\to\UV_1&\arr&\V_1&&&&\D_2\to\UV_2&\arr&\V_2&&&&\D_3\to\\
&\sarr&&\sarr&&\narr&&\sarr&&\sarr&&\narr\\
&&\UCS_1&\arr&\CS_1&&&&\UCS_2&\arr&\CS_2
\end{matrix}
\label{eq:combined}
\end{equation}

\subsection{The last class}
The partial order (\ref{eq:combined}) is clearly missing something, so
we make the following definition.
\begin{equation}
\UD_{n+1}=\P^{\US_n}=\P^{\UCS_n}
\end{equation}

\begin{theorem}
$\US_n\subseteq\UD_{n+1}\subseteq\D_{n+1}$.
\end{theorem}

\proof $\USAT_n\in\P^{\USAT_n}$ and $\US_n\subseteq\S_n$.  \endproof

We add this information to the partial order (\ref{eq:combined}) to
get our final diagram.  Note that the periodic ``bulge'' is a cube.
\begin{equation}
\begin{matrix}
&&\US_1&\arr&\S_1&&&&\US_2&\arr&\S_2\\
&\narr&&\xarr&&\sarr&&\narr&&\xarr&&\sarr\\
\P\to\UV_1&\arr&\V_1&&\UD_2&\arr&
\D_2\to\UV_2&\arr&\V_2&&\UD_3&\arr&\D_3\to\\
&\sarr&&\xarr&&\narr&&\sarr&&\xarr&&\narr\\
&&\UCS_1&\arr&\CS_1&&&&\UCS_2&\arr&\CS_2
\end{matrix}
\label{eq:complete}
\end{equation}

\section{Other properties}
Diagram (\ref{eq:complete}) summarizes our current knowledge about the
partial order in the promise polynomial hierarchy.  However, the
promise polynomial hierarchy is easier to work with than the
non-promise polynomial hierarchy, and there are interesting questions
which we can answer in the promise polynomial hierarchy that are open
in the non-promise polynomial hierarchy.

\subsection{Intersections}
There are two intersections in diagram (\ref{eq:complete}) that we
would like to identify, and it turns out that in the promise
polynomial hierarchy we can do so.  We know that
$\V_n\subseteq\S_n\intersect\CS_n$, and
$\UV_n\subseteq\US_n\intersect\UCS_n$.  The next two theorems show
that these inclusions are, in fact, equalities.

\begin{theorem}\label{thm:Sn*CSn}
$\V_n=\S_n\intersect\CS_n$.
\end{theorem}

\proof We give the proof for $n=1$.

Suppose $\phi\in\S_1\intersect\CS_1$.  Since $\phi\in\S_1=[\SAT]$,
there is a polynomial expression $g(x,y)$ so that $\exists y:g(x,y)=1$
implies $1\in\phi(x)$ and $\forall y,\,g(x,y)=0$ implies
$0\in\phi(x)$.  Since $\phi\in\CS_1=[\CSAT]$, there is also a
polynomial expression $h(x,y)$ so that $\forall y,\,h(x,y)=1$ implies
$1\in\phi(x)$ and $\exists y:h(x,y)=0$ implies $0\in\phi(x)$.

Now define $f_x(i,y)$ (where $|i|=1$) by $f_x(1,y)=g(x,y)$ and
$f_x(0,y)=\lnot h(x,y)$.  By construction, if $f_x(i,y)=1$ then
$i\in\phi(x)$.

Suppose that there is no $(i,y)$ such that $f_x(i,y)=1$.  Then
$\forall y,\,g(x,y)=0$ and $\forall y,\,h(x,y)=1$, so $0\in\phi(x)$
and $1\in\phi(x)$, so $\phi(x)=\A$.

Therefore $\phi(x)\supseteq\VAL(f_x)$, so $\phi\in[\VAL]=\V_1$.
\endproof

\begin{theorem}\label{thm:USn*UCSn}
$\UV_n=\US_n\intersect\UCS_n$.
\end{theorem}

\proof We give the proof for $n=1$.  The proof is similar to the proof
of theorem \ref{thm:Sn*CSn}.

Suppose $\phi\in\US_1\intersect\UCS_1$.  Since $\phi\in\US_1=[\USAT]$,
there is a polynomial expression $g(x,y)$ so that $\exists!y:g(x,y)=1$
implies $1\in\phi(x)$, $\forall y,\,g(x,y)=0$ implies $0\in\phi(x)$,
and if neither of these holds then $\phi(x)=\A$.  Since
$\phi\in\UCS_1=[\UCSAT]$, there is also a polynomial expression
$h(x,y)$ so that $\forall y,\,h(x,y)=1$ implies $1\in\phi(x)$,
$\exists!y:h(x,y)=0$ implies $0\in\phi(x)$, and if neither of these
holds then $\phi(x)=\A$.

Now define $f_x(i,y)$ (where $|i|=1$) by $f_x(1,y)=g(x,y)$ and
$f_x(0,y)=\lnot h(x,y)$.  By construction, $\exists!y:f_x(1,y)=1$
implies $1\in\phi(x)$ and $\exists!y:f_x(0,y)=1$ implies
$0\in\phi(x)$, so $\exists!(i,y):f_x(i,y)=1$ implies $i\in\phi(x)$ for
that value of $i$.

We claim that if it is not true that $\exists!(i,y):f_x(i,y)=1$, then
$\phi(x)=\A$.  Firstly, suppose that there is no $(i,y)$ such that
$f_x(i,y)=1$.  Then $\forall y,\,g(x,y)=0$ and $\forall y,\,h(x,y)=1$,
so $0\in\phi(x)$ and $1\in\phi(x)$, so $\phi(x)=\A$.  Now suppose that
$f_x(i_1,y_1)=f_x(i_0,y_0)=1$ with $(i_1,y_1)\ne(i_0,y_0)$.  We may
suppose that $\phi(x)\ne\A$, so there is at most one $y$ such that
$g(x,y)=1$ and there is at most one $y$ such that $h(x,y)=0$.
Therefore $i_1\ne i_0$, and we may assume $i_1=1$ and $i_0=0$.  Now
$\exists!y:g(x,y)=1$ and $\exists!y:h(x,y)=0$, so $1\in\phi(x)$ and
$0\in\phi(x)$, so $\phi(x)=\A$.

Therefore $\phi(x)\supseteq\UVAL(f_x)$, so $\phi\in[\UVAL]=\UV_1$.
\endproof

\subsection{Weak closures}
There are two weak closures in diagram (\ref{eq:complete}) that we
would like to identify, and it turns out that in the promise
polynomial hierarchy we can do so.  We know that
$\P^{\V_n}\subseteq\P^{\S_n}=\D_{n+1}$, and
$\P^{\UV_n}\subseteq\P^{\US_n}=\UD_{n+1}$.  The next two theorems show
that these inclusions are, in fact, equalities.

\begin{theorem}\label{thm:P^Vn}
$\P^{\V_n}=\D_{n+1}$.
\end{theorem}

\proof We give the proof for $n=1$.  The proof is standard \cite{ESY}.

We show that $\SAT\in\P^{[\VAL]}$, which suffices.  Given a polynomial
expression $f_1(x_1,\ldots,x_m)$, we call the oracle for $\VAL(f_1)$
to get the first bit $x'_1$ of a putative solution.  We then define
$f_2(x_2,\ldots,x_m)=f_1(x'_1,x_2,\ldots,x_m)$ and call $\VAL(f_2)$ to
get the second bit $x'_2$ of a putative solution.  We continue until
we have $x'=(x'_1,\ldots,x'_m)$ and then we evaluate
$f_1(x'_1,\ldots,x'_m)$.  \endproof

\begin{theorem}\label{thm:P^UVn}
$\P^{\UV_n}=\UD_{n+1}$.
\end{theorem}

\proof We give the proof for $n=1$.  The proof is very similar to the
proof of theorem \ref{thm:P^Vn}.

We show that $\USAT\in\P^{[\UVAL]}$, which suffices.  Given a
polynomial expression $f_1(x_1,\ldots,x_m)$ with at most one solution,
we call the oracle for $\UVAL(f_1)$ to get the first bit $x'_1$ of the
putative solution.  We then define
$f_2(x_2,\ldots,x_m)=f_1(x'_1,x_2,\ldots,x_m)$ and call $\UVAL(f_2)$
to get the second bit $x'_2$ of the putative solution.  We continue
until we have $x'=(x'_1,\ldots,x'_m)$ and then we evaluate
$f_1(x'_1,\ldots,x'_m)$.  \endproof

\subsection{Randomized reduction}
We can further weaken weak reduction to random reduction.  If $\phi_1$
and $\phi_2$ are promise problems, we define $\phi_1\propto_R\phi_2$
to mean that there is a $\phi_2$-oracle {\it probabilistic} Turing
machine $M$ and a polynomial $p$ such that, when $M$ is given
$x\in\A^*$, every possible path $\pi$ of $M$ runs in time at most
$p(|x|)$ and computes some $y_\pi$, and $y_\pi\in\phi_1(x)$ with
probability at least $2/3$.  We define
$\BPP^{\phi_2}=\{\phi_1\mid\phi_1\propto_R\phi_2\}$.  If $X=[\phi]$ or
$X=\P^\phi$, we define $\BPP^X=\BPP^\phi$.

We could similarly generalize RP and coRP, which allow a probability
of error for one answer but not the other, but we do not pursue that
possibility here.

Under randomized reduction, it turns out that the entire ``bulge''
from $\UV_n$ to $\D_{n+1}$ is equivalent.  Our proof is a simple
corollary of theorem \ref{thm:P^UVn} and a result of Valiant and
Vazirani \cite{VV}.

\begin{theorem}
$\BPP^{\UV_n}=\BPP^{\D_{n+1}}$.
\end{theorem}

\proof We rely on the result that $\SAT\propto_R\USAT$ \cite{VV,
Goldreich}, and the relativization of this result which states that
$\SAT_n\propto_R\USAT_n$.  Now
$\SAT_n\in\BPP^{\USAT_n}=\BPP^{\UVAL_n}$ by theorem \ref{thm:P^UVn}
and the fact that weak reduction implies random reduction.  \endproof

\section{Bibliography}
\bibliographystyle{plain}
\bibliography{promise}

\end{document}